\documentclass{article}
\usepackage{epsfig}

\renewcommand{\thefootnote}{\fnsymbol{footnote}} 
\def\z{\noindent} 
\def\p{\partial}
\def\ap{\approx}

\begin{document}

\title{Fowler-Nordheim emission modified by laser pulses in the adiabatic regime} 

\author{A.Rokhlenko  and J.L.Lebowitz$^*$\thefootnote
\medskip
\\Department of Mathematics,
Rutgers University\\ Piscataway, NJ 08854-8019}

\maketitle

\begin{abstract}
We investigate enhanced field emission due to a continuous or pulsed oscillating field 
added to a constant electric field $E$ at the emitter surface. When the frequency of 
oscillation, field strength, and property of the emitter material satisfy the Keldysh 
condition $\gamma<1/2$ one can use the adiabatic approximation for treating the 
oscillating field, i.e. consider the tunneling through the instantaneous Fowler-Nordheim 
barrier created by both fields. Due to the great sensitivity of the emission to the 
field strength the average tunneling current can be much larger than the current 
produced by only the constant field. 

We carry out the computations for arbitrary strong constant electric fields, beyond the 
commonly used Fowler-Nordheim approximation which exhibit in particular an important 
property of the wave function inside the potential barrier where it is found to be 
monotonically decreasing without oscillations. 

\end{abstract}
\z 

\footnotetext{$^*$Also Department of Physics} 

\bigskip
In a constant electric field the current due to electron tunneling from a metal is 
described by the commonly used Fowler-Nordheim (FN) equations [1]. They are modified 
to include arbitrary strong fields, see e.g. [2] and [3], though for practical needs 
the low field approximation made in [1] are usually sufficient. The triangular potential 
barrier used in these articles was corrected for image forces by Schottky [4]. They 
are important in constant electric fields, but an adequate method for treating them is 
not clear under laser radiation. Their physical origin is based on the rearrangement 
of the electron distribution inside the emitters; a time dependent process whose 
duration is not known well. To simplify the problem (and losing some precision) the 
Schottky term will not be included here.

In experiments the constant electric fields are often supplemented by short
laser pulses with electric component of amplitude $F$ orthogonal to the metal surface. 
Here we carry out computations for a simple one-dimensional model of the emitting 
surface in order to explore a practically important [5], [6] situation when $F$ can be 
treated adiabatically. This is possible under suitable conditions on the strengths of 
the constant field $E$ and laser field $F$, frequency $\omega$ of the electromagnetic 
oscillations, and the emitter properties. A pulse of duration $T$ is applied at $t=0$. 
The pulse shape is assumed smooth enough and its amplitude $F$ significantly lower than 
$E$. Our goal is to find the current gain in the interval $0<t<T$.

\bigskip\noindent
{\bf Principal approximation}

\medskip
The laser assisted field emission is an important method to increase tunneling current
and have flexible easily manipulated electron sources. The theoretical treatment of this
effect is quite difficult because in involves interplay of different processes, such as 
multi-photon and above barrier emission, FN tunneling, and emitter heating, see for example
[5].

Our simple theory disregards emitter heating, the photofield ionization caused by photon 
absorption from the oscillating field. This can be justified only when the laser field is 
not strong and/or the photon frequency is relatively low for significant contribution of 
multi-photon processes. Nevertheless the optical field changes the shape of the potential 
barrier and therefore [6] the rate of tunneling. Such situation was studied 50 years ago 
by Keldysh in [7] who introduced the parameter $\gamma$ which separates the regions where 
the time of change of the external electric field is longer or shorter than the transition 
time $\tau$ of crossing the potential barrier by electrons, see [7], [8]. The adiabatic 
regime can be realized when the Keldysh parameter $\gamma=\omega \tau <0.5$ where $\omega$ 
is the angular frequency of radiation. In the Keldysh derivation this barrier 
is rectangular. The parameter $\gamma$ has been refined and specified [8] and we modify its 
form taking into account that it is triangular in the FN case. A reasonable expression for 
the time $\tau$ of crossing the potential barrier [9], [10] in the presence of a constant 
field ${\bf E}$ is given by the integral$$ 
\sqrt{m\over 2}\int_0^q{dx\over \sqrt{V-e{\bf E}x}}= {\sqrt{2m}(\sqrt{V}-\sqrt{W})
\over e{\bf E}},\eqno(1)$$
where $V$ is the total height of the potential barrier, $W$ is the location of the
Fermi level at $x<0$, $0$ and $q=(V-W)/eE$ are turning points of tunneling electrons.

Thus the modified Keldysh parameter has the form$$
\gamma = \omega{\sqrt{2m}(\sqrt{V}-\sqrt{W})\over e{\bf E}}.\eqno(2)$$
We assume that $F(t)=0$ for $t<0$, i.e. the electromagnetic radiation starts at $t=0$ when
an electron enters the potential barrier, it goes out of it at $x=q$, therefore $0$ and $q$ are
the limits of integration. When $\gamma <0.5$ the tunneling can be safely approximated as 
an adiabatic process, i.e. by taking for ${\bf E}$ in (2) its instantaneous value of the time 
dependent sum ${\bf E}(t)=E+F(t)$. $q=V/eE$ is evaluated by Eq(8A) in Appendix for a constant 
$E$.

Disregarding the relatively small $F$ (compared with $E$) we finally consider here only the 
regime when$$
\omega < {eE\over 2\sqrt{2m}(\sqrt{V}-\sqrt{W})}.\eqno(3)$$
As an illustration we show the results assuming $V=2W$ and study two cases for the work
function $\chi=5$, $\chi=2.1$ eV [9]. By denoting $F/E=\beta<1$ and treating $\beta$ as a small 
quantity Eq.(3) can be presented in two forms$$
\omega < p\cdot 10^{14}E,\ {\rm or}\ \lambda>{h\cdot 10^3\over E},\eqno(4)$$
where $\omega$, is in $sec^{-1}$, $\lambda$ in nm, and $E$ in V/nm. Shown in Fig.1 are two 
curves of laser wavelengths $\lambda(E)$ ($\beta =0$ in Fig.4 for simplification). For $\lambda$
above the curves the adiabatic approximation is valid. The parameters are $p\ap 1.6$, 
$h\ap 1.18$ for Au, Cu, Ni, W, and $p\ap 2.5$, $h\ap 0.75$ for Cs.
\vskip0.6cm
\hskip1.7cm
\epsfig{file=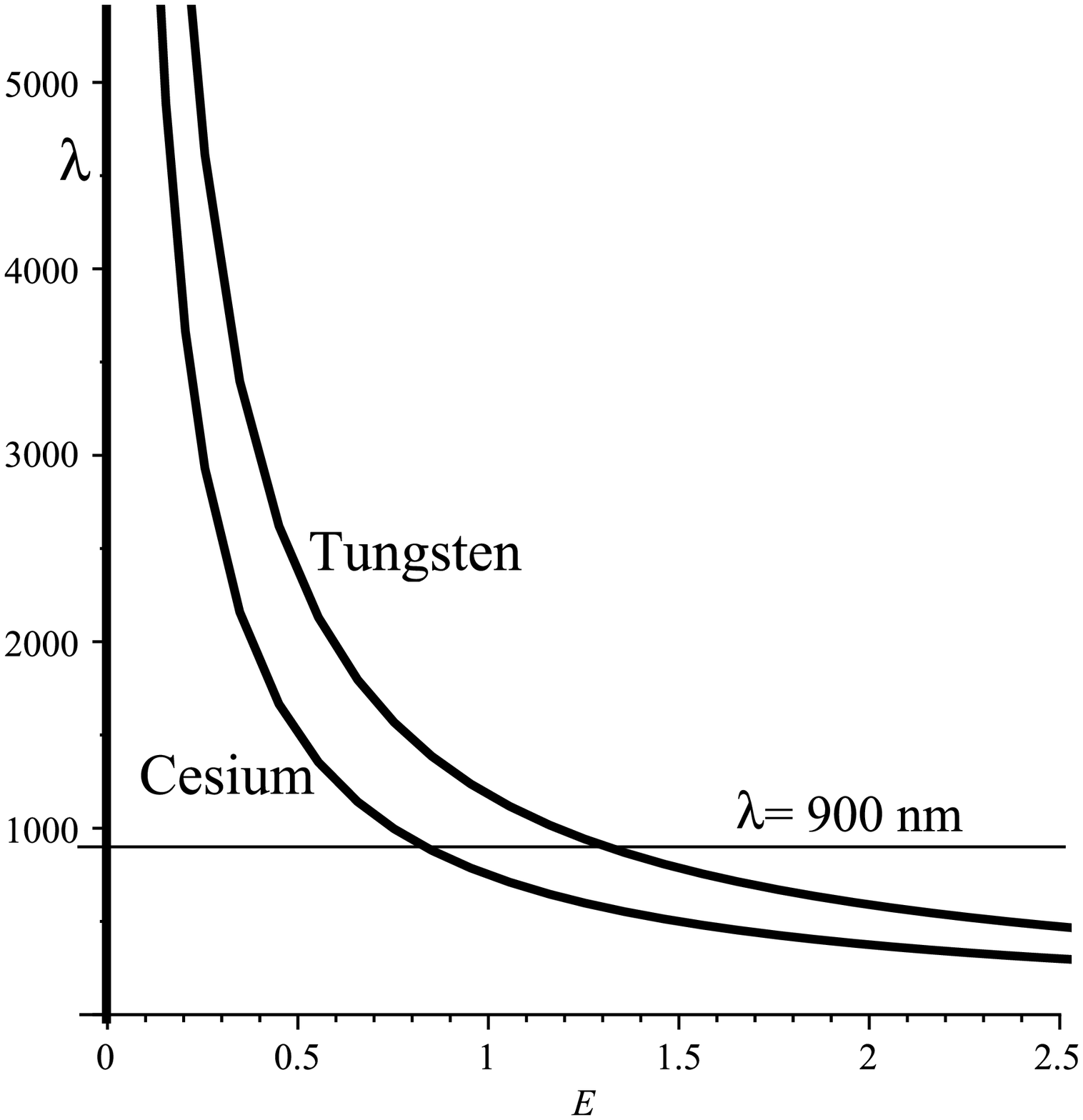, width=8cm, height=7cm}

\centerline 
{\small FIG.1. Laser wavelengths $\lambda$ above corresponding curves can be treated 
adiabatically}
\vskip0.3cm\noindent
Results in Fig.1 are approximate because the work functions are not the same for different 
crystal orientations. Nevertheless Fig.1 gives a general picture of permissible regions for 
treating tunneling by the adiabatic method when emitters are irradiated by lasers of frequencies 
estimated in our work.

The wavelength of short laser pulses [10], [11], [12] frequently used in experiments and theory 
is around $\lambda=800-900$ nm, see Fig.1. Note also that in strong fields $E$ about $50-100$ 
V/nm the FN approximate Eq.(9A) for the tunneling probability $P$ becomes incorrect [2].

\bigskip\noindent
\noindent{\bf Results of calculations}

\medskip
\noindent
1. Rectangular laser pulse

\medskip
On the interval $0<t<T$ the total field has the form$$
{\bf E}(t)=E+F \cos(\omega t),\ \ E,F>0,\ \ 0<t<T,\eqno(5)$$
we simplify our computations by evaluating the tunneling probability $P(t)=P({\bf E}(t))$ and 
treating ${\bf E}(t)$ as a fixed field in (5) in spirit of adiabatic approximation. Then we 
evaluate $\tilde {P}$ as the average of $P(t)$ on $(0,T)$ when both fields $E$ and $F(t)$ are 
turned on$$ 
\tilde{P}={\omega\over 2\pi}\int_0^{2\pi/\omega}{P(t)dt}.\eqno(6)$$
We use here the Eq.(5) which allows to compute $\tilde{P}$ on a shorter time interval of a 
single period of laser field when the pulse is rectangular.
The entries in Table 1 are $\tilde{P}/P_E$, where $P_E$ is the exact tunneling probability in 
the constant field $E$ given in Eq.(8A). Note that $E$ is equal to the time average of 
${\bf E}(t)$. The gain of electron tunneling flow during the pulse period is multiplied by the 
number of oscillations in a single pulse which will correspond to the rectangular laser pulse 
of duration $T$. In the case of other pulse shapes such calculation is longer because it should 
be extended on the whole pulse duration.

Fig.3 in Appendix implies that for the adiabatic treatment we always are quite far from the 
maximum point of $P(t)$, i.e. the same results can be found using Fowler-Nordheim formula (9A). 
In Table 1 are shown the results for Tungsten and Cesium in the electric fields permissible for
the adiabatic regimes. The entries in Table 1 exhibit the average current gain in time of the 
pulse action compared with the case when $\beta$ and $F$ are zero. Note that we use everywhere 
in this work the CGS unit system.
\vskip0.3cm
\centerline {TABLE 1}
\vskip0.3cm
\hskip1.0cm 
\begin{tabular}{l*{6}{c}r}

\hline
\hline
Tungsten       &$E=1.5$   &$E=2.0$ &$E=3$   &$E=5$    &$E=8$ \\
\hline
$\beta=0.1$     & 20.50    & 7.468  & 2.917  & 1.542  & 1.183 \\
$\beta=0.2$     & 745.7    & 102.6  & 14.99  & 3.550  & 1.750 \\ 
\hline
\hline
Cesium         &$E=1.0$   &$E=1.5$ &$E=2.0$  &$E=3.0$ &$E=5.0$ \\
\hline
$\beta=0.1$     & 2.137    & 1.432  & 1.223  & 1.086  & 1.021 \\
$\beta=0.2$     & 7.612    & 2.948  & 1.926  & 1.341  & 1.083 \\
\hline
\hline
\end{tabular}
\vskip0.5cm

One can see that the lower is the constant background electric field the stronger is the 
effect of laser radiation caused by its negative half-waves. This can be easily justified 
by using Eq.(9A) for weaker electric fields and evaluating the ratio$$
{1\over P_{FN}} {d P_{FN}\over dE}={4\sqrt{2m}\over 3eE^2\hbar}\chi^{3/2},\eqno(7)$$
which grows rapidly when $E$ is relatively small and decreasing. This property depends  
strongly on the shape of the pulses.

\medskip
\noindent
2. Laser pulse with envelopes $sin(t\pi/T)$ and triangular one

\medskip
We consider first the laser pulse $F\sin(t\pi/T)\cos(\omega t)$ for $0<t<T$ where 
$T=16\pi/\omega$, which means that the pulse consists of 8 periods. As the average pulse 
amplitude is lower than of the rectangular one the gain is smaller too but substantial 
anyway when $E$ is low. Table 2 below is similar to Table 1 and in both of them the 
tunneling current grows from $10^{-22}$ to $10^{-4}$ for Tungsten and from $10^{-9}$ to 
$0.03$ for Cesium when $E$ increases, but the laser pulse effect gets smaller.
\vskip0.3cm
\centerline {TABLE 2}
\vskip0.3cm
\hskip1.0cm 
\begin{tabular}{l*{6}{c}r}

\hline
\hline
Tungsten       &$E=1.5$   &$E=2.0$ &$E=3$   &$E=5$    &$E=8$ \\
\hline
$\beta=0.1$     & 8.733    & 3.811  & 1.899  & 1.263  & 1.089 \\
$\beta=0.2$     & 230.3    & 37.13  & 6.871  & 2.204  & 1.367 \\ 
\hline
\hline
Cesium         &$E=1.0$   &$E=1.5$ &$E=2.0$  &$E=3.0$ &$E=5.0$ \\
\hline
$\beta=0.1$     & 1.544    & 1.210  & 1.108  & 1.041  & 1.021 \\
$\beta=0.2$     & 3.943    & 1.931  & 1.452  & 1.167  & 1.010 \\
\hline
\hline
\end{tabular}
\vskip0.3cm
The shape of laser pulse envelopes is often close to a triangle. In Table 3 we show the
results for isosceles triangular pulses symmetric about 
the vertical axis of the same maximum amplitudes as above only for a Tungsten emitter.
\vskip0.3cm
\centerline {TABLE 3}
\vskip0.3cm
\hskip1.0cm 
\begin{tabular}{l*{6}{c}r}

\hline
\hline
Tungsten       &$E=1.5$   &$E=2.0$ &$E=3$   &$E=5$    &$E=8$ \\
\hline
$\beta=0.1$     & 5.685    & 2.785  & 1.594  & 1.178  & 1.061 \\
$\beta=0.2$     & 125.2    & 21.69  & 4.652  & 1.797  & 1.249 \\ 
\hline
\hline
\end{tabular}

\vskip0.5cm\noindent
Similar results can be easily obtained for any envelope of electromagnetic pulses.

The electric field amplitude $F_0$ of the laser radiation for computations [13] can be 
evaluated by using the Pointing vector$$
\vec S=\vec E\times \vec H= {1\over 2} c\epsilon_0 E^2_0\vec k,\eqno(8)$$
where $c$ is the speed of light, $\epsilon_0$ - vacuum permittivity, and $\vec k$ gives the 
direction of the beam with rectangular cross section. Using the well known constants the
amplitude of the field $E_0$ can be presented as
$$E_0(V/m)=27.42\sqrt{{Beam\ power\ (watt)\over Beam\ cros section\ (m^2)}}.\eqno(9)$$  
Here units are given in round parentheses. After adding the constant field this equation 
allows to find the optical field.

When $\gamma>10$ the main contribution to electron emission comes from multiphoton processes
and emitter heating [10],[11], the case $1<\gamma<10$ is more difficult. Under conditions 
given here one can use the adiabatic technique, which in our work is implemented to show that 
the electromagnetic radiation can very significantly increase the emission, see Tables 1-3. 

\bigskip\centerline
{\bf APPENDIX}
\bigskip

We supplement the results for evaluating the time independent tunneling in an arbitrary strong 
electric field derived in [1] with some additional details. Let us consider a metallic block, 
which is placed to the left of $x=0$, an electron with kinetic energy corresponding the Fermi 
level $W=\hbar^2k^2/2m$ enters the triangular potential field whose shape is determined by the 
constant electric fields $E$ on $-\infty<x<\infty$ and $-Fx$ on the beam $x>0$.
The governing Schr\"{o}dinger equation for the electron wave function $\psi (x,t)$ 
on the infinite interval $-\infty <x<\infty$ can be written in the following form
$$i\hbar{\p \psi\over\p t}(x,t)=-{\hbar^2\over 2m}{\p^2\psi\over\p x^2}(x,t)+(V-
eEx - eFx\cos{\omega t})\psi(x,t),\ E,F>0.\ \ \eqno(1A)$$
We neglect here the Schottky term, here $e,\ m$ are the electron mass and charge, $F$ is 
the amplitude of the dipole field, which assumed to be orthogonal to the emitter surface. 
Heating of a bulk emitter is neglected here. 

The tunneling is described by the stationary equation$$
-{\hbar^2\over 2m}{\p^2\psi\over\p x^2}(x)+(V-eEx)\psi(x)=W\psi(x),\eqno(2A)$$
whose exact solution in the Bessel functions, found in [1], can be written in the 
form$$
\psi (x)= \left\{\begin{array}{rcl}
\exp (ikx)+a_1\exp (-ikx) &\mbox{for}& x<0,\\
a_2\sqrt{y}I_{1/3}(2y^{3/2}/3)+a_3\sqrt{y}K_{1/3}(2y^{3/2}/3) &\mbox{for}&
 0<x<q,\\ 
a_4\sqrt{z}H^{(1)}_{1/3}(2z^{3/2}/3) &\mbox{for}& x>q.\\
\end{array}\right\}\eqno(3A)$$
where $q=(V-W)/eE.$           
The variables $y$ and $z$ are non-negative in their domains and they are$$ 
y=(V-W-eEx)(2m)^{1/3}(eE\hbar)^{-2/3},\ \ 
z=(eEx+W-V)(2m)^{1/3}(eE\hbar)^{-2/3}.\eqno(4A)$$

Inside the emitter the tunneling electrons with the Fermi energy $W=(\hbar k)^2/2m$ and wave 
their function it is represented by the first term in Eq.(3), where $e^{ikx}$ creates the 
incoming current $j_{-}=\hbar k/m$ toward the surface from its left side.
The coefficients $a_1,...,a_4$, calculated in [2] using the continuity of $\psi(x)$
and its derivatives at $x=0$ and $x=q$ are:
$$a_{2}=-{2a_4e^{\pi i/3}},\ \ a_3={2a_4\over \pi i},$$
\vskip-0.5cm
$$\eqno(5A)$$
\vskip-0.5cm
$$a_1=a_4\rho^{1/3}\left\{-e^{\pi i\over 3}I_{1/3}(\rho)-iK_{1/3}(\rho)/\pi+i\sigma
\left[e^{\pi i\over 3}I_{2/3}(\rho)+{e^{\pi i\over 6}\over \pi}K_{2/3}(\rho)\right]\right\},$$
where$$
a_4={-2(2/3)^{1/3}\rho^{-1/3}\over {e^{\pi i/3}I_{1/3}(\rho)+iK_{1/3}(\rho)/\pi+
i\sigma e^{\pi i/3}[I_{2/3}(\rho)+e^{-\pi i/6}K_{2/3}(\rho)/\pi]}}. \eqno(6A)$$
Here $$
\rho ={2\sqrt{2m}\chi^{3/2}\over3eE\hbar},\ \ \ \sigma=\sqrt{\chi\over W},\ \ \ \chi =V-W.
\eqno(7A)$$
The transversal time of crossing the potential barrier was estimated in [5], [6] and
used below, $\chi$ is the work function. We emphasize here that the wave function inside 
the barrier where $0\leq y\leq q$ is monotonically decreasing without oscillations, as 
this can be seen in Fig.2.

\vskip0.3cm
\hskip1.7cm
\epsfig{file=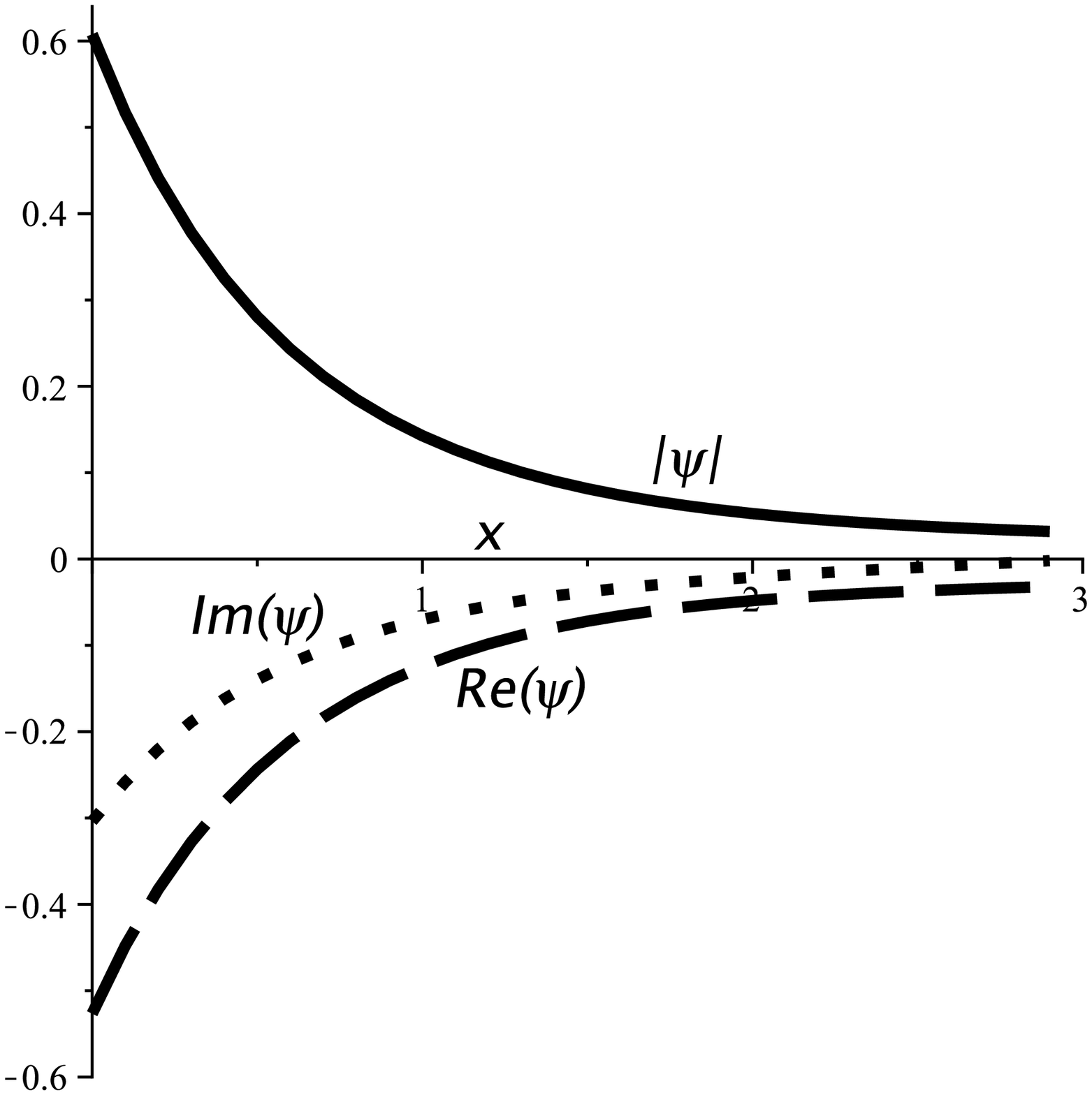, width=8cm, height=6.5cm}

\centerline 
{\small FIG. 2. Wave function $\psi(x)$ inside the barrier}
\medskip
\noindent
Here the $x$-scale given in units proportional to the real one, the amplitude of $\psi$ 
in Fig.2 should be multiplied by $40 a_4$ to correspond Eq.(3). The general behavior of
curves corresponding different parameters $\sigma,\ \rho$ is similar to the ones in Fig.2.

The Hankel function $H^{(1)}$ allows to find the moving to the right current $j_{+}$ 
of escaped electrons proportional to $|a_4|^2$. Thus using Eqs.(5-7) the stationary
tunneling probability for $t<0$ when $F=0$ is$$
P={j_{+}\over j_{-}}={2\sigma\over \pi\rho\big |e^{\pi i\over 3}I_{1/3}(\rho)+
iK_{1/3}(\rho)/\pi +\sigma [e^{5\pi i\over 6}I_{2/3}(\rho)+e^{2\pi i\over 3} 
K_{2/3}(\rho)/\pi]\big |^2}.\eqno(8A)$$  

In practice the usual approximate equation, derived in [1] almost a century ago, with 
a constant electric field, $$
P_{FN}={4\sqrt{W\chi}\over V}\exp{\left(-{4\sqrt{2m}\over 3eE\hbar}
\chi^{3/2}\right)}={4e^{-2\rho}\over \sigma+\sigma^{-1}},\eqno(9A)$$
provides the same results for $P$ as the exact Eq.(8) for $\rho^{1/3}>1$, which
corresponds to all $E<60$ V/nm for Copper with $\chi\ap 5$ eV and not very different
for other metals (but Cs, where $\chi =2.1$). It was shown in [1] that in very strong 
fields $E$ the tunneling probability $P$ is very close to $P_S=5.44 \sigma^{-1}\rho^{1/3}$ 
which helps to construct an interpolation equation$$
P_{int}={1\over 1/P_{FN}+1/P_S}.\eqno(10A)$$
The results provided by Eq.(10A) give decent approximations [1] with a slightly higher 
maximum than the exact Eq.(8A), including the location of the maximum of $P$ in a wide 
range of problem parameters.

Below are shown the exact tunneling probability $P$ and its approximations $P_S$, $P_{FN}$ 
as functions of the dimensionless parameter $r=\rho^{1/3}$ in a case when $\sigma =1$, 
i.e. $V=2W$. These approximate functions become unphysical in the region around the
maximum of $P$, which they cannot describe. 
\vskip0cm
\hskip1.7cm
\epsfig{file=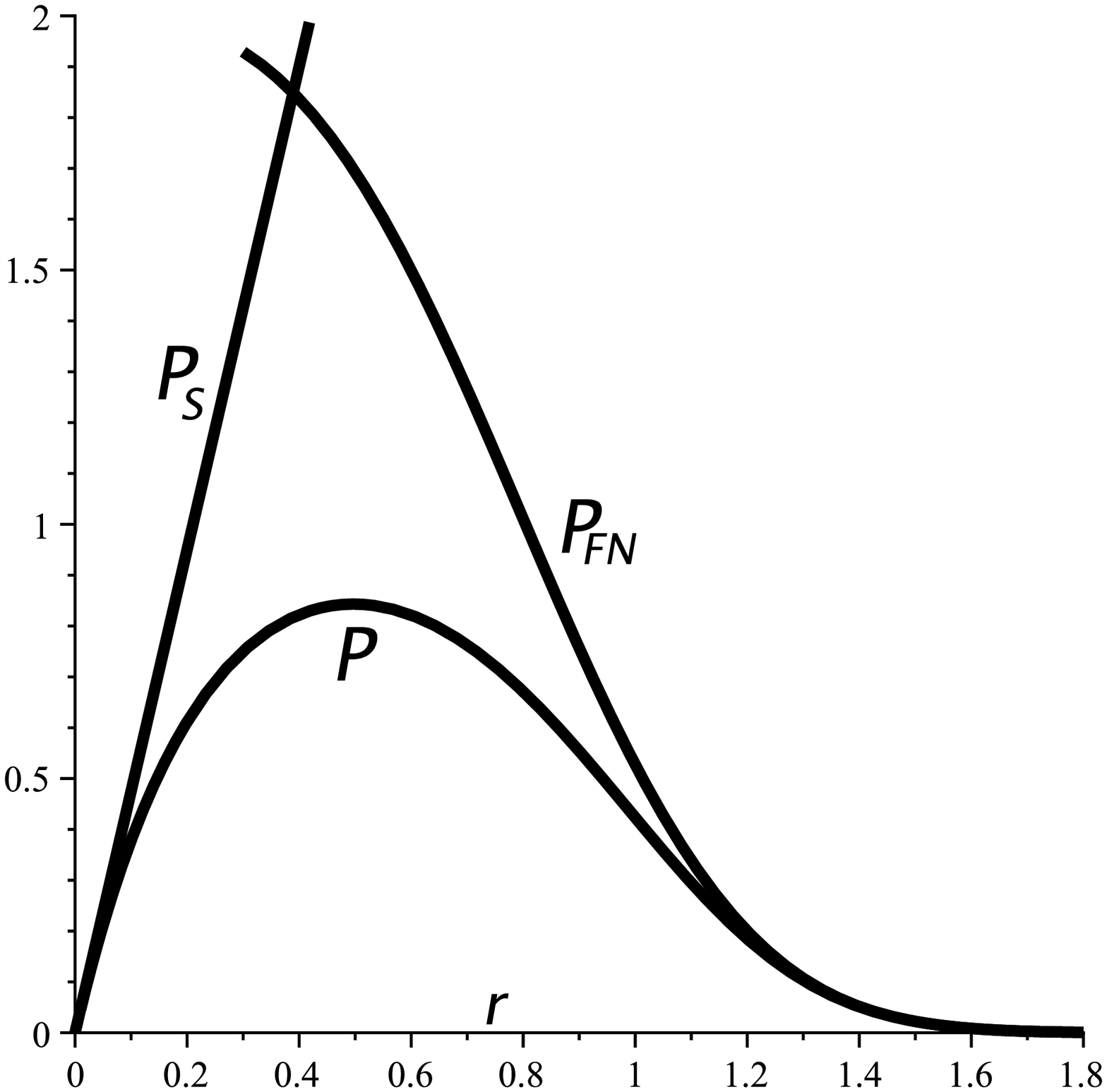, width=8cm, height=5cm}

\centerline 
{\small FIG. 3. Tunneling probability $P$ and its approximations}

\medskip\noindent
Parameter $r$ depends on $E$ and the work function $\chi$, therefore the horizontal 
scale corresponds to different electric fields for different materials.

\bigskip
{\it Acknowledgments} 

\medskip
We thank Kevin Jensen and Don Shiffler for valuable discussions.

Research supported by AFOSR Grant F49620-01-0154.

\end{document}